\documentclass[aps,prd,floatfix,superscriptaddress,showpacs,showkeys]{revtex4}

\usepackage{graphicx,epsfig,epstopdf}
\usepackage{amssymb,amsmath,amsxtra,amsfonts}
\usepackage{bm}

\usepackage{longtable}
\usepackage{multirow}
\usepackage{booktabs}
\usepackage{array}
\usepackage{wrapfig}
\usepackage{color}

\usepackage{titlesec}
\titleformat{\section}{\large\bfseries}{\thesection}{1em}{}

\newcommand{\bea}{\begin{eqnarray}}
\newcommand{\ena}{\end{eqnarray}}
\newcommand{\be}{\begin{equation}}
\newcommand{\en}{\end{equation}}
\newcommand{\nn}{\nonumber\\}
\newcommand{\ed}{\end{document}} 
\newcommand{\Tr}{\mbox{\rm{tr}}}

\newcommand{\Jpsi}{\ensuremath{J\!/\!\psi}}

\begin{document}

\hfill MITP/16-085 (Mainz)

\title{Four-quark structure of the $Z_c(3900)$, 
$Z(4430)$, and $X_b(5568)$ states}

\author{Fabian~Goerke}
\affiliation{Institut f\"ur Theoretische Physik, Universit\"at T\"ubingen,
Kepler Center for Astro and Particle Physics,
Auf der Morgenstelle 14, D-72076, T\"ubingen, Germany}

\author{Thomas~Gutsche}
\affiliation{Institut f\"ur Theoretische Physik, Universit\"at T\"ubingen,
Kepler Center for Astro and Particle Physics,
Auf der Morgenstelle 14, D-72076, T\"ubingen, Germany}

\author{Mikhail~A.~Ivanov}
\affiliation{Bogoliubov Laboratory of Theoretical Physics, 
Joint Institute for Nuclear Research, 141980 Dubna, Russia}

\author{J\"urgen~G.~K\"orner}
\affiliation{PRISMA Cluster of Excellence, Institut f\"{u}r Physik, 
Thomas~GutscheJohannes Gutenberg-Universit\"{a}t,  
D-55099 Mainz, Germany}

\author{Valery~E.~Lyubovitskij}
\affiliation{Institut f\"ur Theoretische Physik, Universit\"at T\"ubingen,
Kepler Center for Astro and Particle Physics,
Auf der Morgenstelle 14, D-72076, T\"ubingen, Germany}
\affiliation{ 
Department of Physics, Tomsk State University,  
634050 Tomsk, Russia} 
\affiliation{Laboratory of Particle Physics, 
Mathematical Physics Department, 
Tomsk Polytechnic University, 
Lenin Avenue 30, 634050 Tomsk, Russia} 

\author{Pietro~Santorelli}
\affiliation{Dipartimento di Fisica, Universit\`a di Napoli
Federico II, Complesso Universitario di Monte Sant'Angelo,
Via Cintia, Edificio 6, 80126 Napoli, Italy}
\affiliation{Istituto Nazionale di Fisica Nucleare, Sezione di
Napoli, 80126 Napoli, Italy}

\begin{abstract}
We examine the four-quark structure of the recently discovered
charged $Z_c(3900)$, $Z(4430)$, and $X_b(5568)$ states. We calculate the widths
of the strong decays $Z_c^+ \to \Jpsi\pi^+$ ($\eta_c\rho^+$,
$\bar D^0D^{\ast\,+}$, $\bar D^{\ast\,0}D^+$),
$Z(4430)^+ \to J/\psi \pi^+$ ($\psi(2s) \pi^+$), 
and $X^+_b\to B_s\pi^+$ within a covariant quark model previously
developed by us.
We find that the tetraquark-type current widely used in the literature
for the $Z_c(3900)$ leads to a significant suppression of the $\bar D D^\ast$
and $\bar D^\ast D$ modes. Contrary to this a molecular-type current provides
an enhancement by a factor of 6-7 for the $\bar D D^\ast$ modes compared with
the $Z_c^+\to\Jpsi\pi^+$, $\eta_c\rho^+$ modes in agreement with
recent experimental data from the BESIII Collaboration. 
In the case of the $Z(4430)$ state we test a sensitivity of the ratio $R_Z$ 
of the $Z(4430)^+ \to \psi(2s) \pi^+$ and $Z(4430)^+ \to J/\psi \pi^+$ 
decay rates to a choice of the size parameter $\Lambda_{Z(4430)}$ 
of the $Z(4430)$. 
Using the upper constraint for the sum of these two modes deduced from 
the LHCb Collaboration data we find that $R_Z$ varies from 4.64 
to 4.08 when $\Lambda_{Z(4430)}$ changes from 2.2 to 3.2 GeV. 
Also we make the prediction for the 
$Z(4430)^+ \to D^{\ast\,+} \bar D^{\ast\,0}$ decay rate. 



\end{abstract}

\maketitle

\section{Introduction}
\label{sec:intro}

In the course of experimentally establishing the heavy meson spectrum unusual 
states were observed that cannot be simply interpreted in the context of  
a minimal constituent quark-antiquark model.  
Among these new states are the $Z_c(3900)$, $Z(4430)$, and the $X_b (5568)$,  
where especially the last resonance still needs solid experimental   
confirmation. The flavor structure of these states is unusual  
as evident from their strong decay modes; a simple quark-antiquark  
interpretation is not feasible.  
In the following we focus on these special states. We first collect 
the experimental findings. 


The process $e^+e^-\to\pi^+\pi^-\Jpsi$ has been studied 
by the BESIII Collaboration \cite{Ablikim:2013mio}.
A structure was observed at around 3.9~GeV in the $\pi^\pm\Jpsi$ mass 
spectrum that was christened the $Z_c(3900)$ state. If interpreted as a new 
particle, it is  unusual in that it carries an electric charge and couples
to charmonium. A fit to the $\pi^\pm\Jpsi$ invariant mass spectrum
results in a mass of 
$M_{Z_c} = 3899.0 \pm 3.6({\rm stat}) \pm 4.9({\rm syst})$~MeV 
and a width  
of $\Gamma_{Z_c}=46 \pm 10({\rm stat}) \pm 20({\rm syst})$~MeV. 


The cross section for $e^+ e^- \to \pi^+ \pi^- J/\psi$ between 3.8 and 
5.5~GeV was measured by the Belle Collaboration~\cite{Liu:2013dau}. 
This measurement lead to the observation of the state $Y(4260)$, and its
resonance parameters were determined. 
In addition, an excess of $\pi^+ \pi^- J/\psi$ production around 4~GeV 
was observed. This feature can be described by a Breit-Wigner parametrization 
with properties that are consistent with the $Y(4008)$ state that was 
previously 
reported by Belle. In a study of the $Y(4260) \to \pi^+ \pi^- J/\psi$ decays, 
a structure was observed in the $M(\pi^\pm\Jpsi)$ mass spectrum with 
$5.2~\sigma$ significance, with mass 
$M=3894.5\pm 6.6({\rm stat})\pm 4.5({\rm syst})$~MeV 
and width $\Gamma=63\pm 24({\rm stat})\pm 26({\rm syst})$~MeV, 
where the errors are statistical 
and systematic, respectively. This structure can be interpreted as 
a new charged charmoniumlike state. 


Using 586~pb of $e^+e^-$~annihilation data the CLEO-c detector made an analysis 
at $\sqrt{s}=4170$~MeV at the peak of the charmonium resonance $\psi(4160)$. 
The subsequent decay $\psi(4160)\to \pi^+\pi^-\Jpsi$ was analyzed
\cite{Xiao:2013iha}, 
and the charged state $Z_c^\pm(3900)$ was observed, which decays into 
$\pi^\pm\Jpsi$ at a significance level of $>5\,\sigma$. 
The value of the mass 
$M_{Z_c} = 3886 \pm 4 ({\rm stat}) \pm 2 ({\rm syst})$~MeV and the width 
$\Gamma_{Z_c}= 37 \pm 4 ({\rm stat}) \pm 8 ({\rm syst})$~MeV were found to be in 
good agreement with the results for this resonance reported by the BES III 
and Belle collaborations in the decay of the resonance $Y(4260)$. 
In addition CLEO-c presented the first evidence for the production of 
the neutral member of  this isospin triplet, $Z_c^0(3900)$ decaying into 
$\pi^0\Jpsi$ at a $3.5\,\sigma$ significance level. 


A study of the process $e^+e^-\to\pi^\pm (D\bar D^\ast)^\mp$
was reported by the BESIII Collaboration~\cite{Ablikim:2013xfr}
at $\sqrt{s}=4.26$~GeV using a 525~pb$^{-1}$ data sample collected with 
the BESIII detector at the BEPCII storage ring. A distinct charged structure 
was observed in the $ (D\bar D^\ast)^\mp $ invariant mass distribution. 
When fitted to a mass-dependent-width Breit-Wigner line shape, the pole mass 
and width were determined to be 
$M_{\rm pole} =  3883.9 \pm 1.5 ({\rm stat}) \pm 4.2({\rm syst})$~MeV and 
$\Gamma_{\rm pole}  = 24.8 \pm 3.3 ({\rm stat}) \pm 11.0({\rm syst})$~MeV. 
The mass and width of the structure referred to as $Z_c(3885)$ 
are $2 \sigma$ and $1 \sigma$, respectively, below those of 
the $Z_c(3900)\to \pi^\pm\Jpsi$ peak observed by BESIII and Belle in 
$\pi^+\pi^-\Jpsi$ final states produced at the same center-of-mass energy. 
The angular distribution of the $\pi Z_c(3885)$  system favors 
a $J^P = 1^+$ quantum number assignment for the structure and 
disfavors the assignment $1^-$ or $0^-$. The Born cross section times 
the $DD^\ast$ branching fraction of the $Z_c(3885)$ is measured to be 
\bea 
\sigma\left(e^+e^-\to \pi^\pm Z_c^\mp(3885)\right)
\times {\cal B}\left( Z_c^\mp(3885)\to (D\bar D^\ast)^\mp\right)  
= 83.5 \pm 6.6 ({\rm stat}) \pm 22.0 ( {\rm syst})~{\rm pb}\,. 
\ena

Assuming that the $Z_c(3885)\to D\bar D^\ast$ signal reported in 
\cite{Ablikim:2013xfr} and  the $Z_c(3900)\to \pi\Jpsi$ signal are 
from the same source, the ratio of partial widths is determined as 
\bea 
\frac{\Gamma(Z_c(3885)\to D\bar D^\ast)}{\Gamma(Z_c(3885)\to \pi\Jpsi)}
= 6.2 \pm 1.1 ({\rm stat}) \pm 2.7({\rm syst})\,. 
\ena
That means that the $Z_c(3900)$ state has a much stronger coupling to $DD^\ast$
than to $\pi\Jpsi$ \cite{Liu:2015pma}. An unbinned maximum likelihood fit gives
a mass of $M=3889.1 \pm 1.8$~MeV and a width of $\Gamma = 28.1\pm 4.1$~MeV
($M=3891.8 \pm 1.8$~MeV and $\Gamma = 27.8\pm 3.9$~MeV)
for the two data sets, respectively. The pole position of this peak 
is calculated to be $M_{\rm pole} = 3883.9 \pm 1.5 \pm 4.2$~MeV
and  $\Gamma_{\rm pole} = 24.8 \pm 3.3 \pm 11.0$~MeV.
The mass and width of the peak observed in the $DD^\ast$ final state agree
with that of the $Z_c(3900)$. Thus, they are quite probably the same state.


The charmoniumlike structure  $Z_c^+(3900)$ was identified
in Ref.~\cite{Dias:2013xfa} as the charged partner of the $X(3872)$ state. 
The  X(3872) meson is considered to be a four-quark state with quantum numbers
$I^G(J^{PC}) = 0^+ (1^{++})$.
The $Z_c(3900)$ meson is interpreted as the isospin 1 partner of the $X(3872)$.
As in Ref.~\cite{Faccini:2013lda} it was assumed that the quantum numbers for 
the neutral state in the isospin multiplet were $I^G(J^{PC}) = 1^+ (1^{+-})$
Using standard QCD sum rules techniques, the coupling constants of 
the $Z^+_c\Jpsi\pi^+$, $Z_c^+\eta_c\rho^+$, $Z_c^+\bar D^0 D^{\ast\,+}$, and  
$Z_c^+\bar D^{\ast\,0} D^+$ vertices 
and the corresponding decay widths were calculated  
with the following results: 
\bea
\Gamma(Z^+_c\to\Jpsi+\pi^+) &=& (29.1 \pm 8.2)\,\text{MeV},
\nn
\Gamma(Z^+_c\to\eta_c+\rho^+) &=& (27.5 \pm 8.5)\,\text{MeV},
\nn
\Gamma(Z^+_c\to \bar D^0 + D^{\ast\, +}) &=& (3.2 \pm 0.7)\,\text{MeV},
\nn
\Gamma(Z^+_c\to \bar D^{\ast\, 0} + D^+) &=& (3.2 \pm 0.7)\,\text{MeV}.
\label{eq:Navarra}
\ena 


The observation of a narrow structure, $X(5568)$, in the decay 
sequence $X(5568) \rightarrow B_s^0 \pi^{\pm}$, 
$B_s^0 \rightarrow J/\psi \phi$, 
$J/\psi\rightarrow \mu^+ \mu^-$, $\phi \rightarrow K^+K^-$ was reported
in \cite{D0:2016mwd} by the D0 Collaboration.
This would be the first  observation of a hadronic state 
with valence quarks of four different flavors. 
The mass and width of the new state are measured to be 
$M = 5567.8 \pm 2.9 {\rm \thinspace (stat)}^{+0.9}_{-1.9} 
{\rm \thinspace (syst)}$~MeV/$c^2$, and 
$\Gamma = 21.9 \pm 6.4 {\rm \thinspace (stat)} ^{+5.0}_{-2.5} 
{\rm \thinspace (syst)}$~MeV/$c^2$. However, in recent analysis 
performed by the LHCb Collaboration an existence of the claimed 
$X(5568)$ state has been not confirmed~\cite{Aaij:2016iev}. 

The observed strong decay mode of the $X^\pm(5586)$ implies flavor structures 
of the type $X^+(s \bar b u \bar d)$.
Since  the $B_s^0 \pi^+$ pair is  produced  in an S wave, 
its quantum  numbers  would be $J^P = 0^+$.
As already pointed out in Ref.~\cite{D0:2016mwd}  
the significant difference between the mass of the $X(5568)$ and 
$M_{B} + M_K$ threshold does not favor a hadronic 
molecular interpretation of the $X(5568)$. 
First qualitative considerations point to the $X(5568)$ state 
being a tetraquark state. 

Structure issues of the $X(5568)$ have already been discussed 
in a number of theoretical
papers~\cite{Agaev:2016mjb}-\cite{Albaladejo:2016eps} 
suggesting various tests both for the tetraquark and hadronic molecular
structure of the $X(5586)$ state. In Ref.~\cite{Burns:2016gvy} 
a few options for the interpretation of the $X(5586)$ state have been 
checked. It was concluded that threshold, cusp, molecular, and tetraquark 
approaches for the explanation of the $X(5586)$ state are all disfavored.
One of the important conclusions was that the mass of the $(bsqq)$ tetraquark 
state must be heavier than the $\Xi_b(5800)$ baryon. Also the authors 
of Ref.~\cite{Burns:2016gvy} deduced a lower limit for the masses of  
a possible $(bsud)$ tetraquark state: 6019 (6107) MeV.  
Complementary to Ref.~\cite{Burns:2016gvy} Ref.~\cite{Guo:2016nhb}
presented an analysis based on general properties of QCD to analyze the
$X(5568)$ states. 
In particular, it was shown that the mass of the $(bsud)$ tetraquark state 
must be bigger than the sum of the masses of the $B_s$ meson and the
light quark-antiquark resonance leading to an estimate of the lower limit 
of $M_{bsud} \sim 5.9$ GeV. Reference~\cite{Esposito:2016itg} 
used a $B_s\pi$-$B\bar{K}$ coupled channel analysis with 
an interaction derived from heavy hadron chiral perturbation theory 
to implement the unitarity feature of the spectrum reported by the
D0 Collaboration. The analysis lead to a  $T$-matrix momentum cutoff 
of $\Lambda = 2.80 \pm 0.04$ GeV,  
which is much larger than a typical scale $\Lambda \simeq 1$GeV. 

Reference~\cite{Ali:2016gdg} estimated the mass of the  lightest $(bsud)$, $0^+$
tetraquark in the framework of a tightly bound diquark model. 
Their semiquantitative analysis leads to a mass of about 5770 MeV 
that lies approximately 200 MeV above 
the reported $X(5568)$ state, and 7 MeV below the $B\bar K$ threshold. 


The $Z(4430)$ state with mass $M = 4433 \pm 4 \pm 2$ MeV and width
$\Gamma = 45^{+18}_{-13} {\rm \thinspace (stat)}^{+30}_{-13}{\rm \thinspace (syst)}$ MeV
has been discovered by the {\it BABAR} Collaboration
in the $\pi^\pm \psi(2s)$ invariant mass distribution in
$B \to K \pi^\pm \psi(2s)$ decay~\cite{4430belle1},
where $\psi(2s)$ is the
first radial excitation of the $J/\psi$.
Later, in Ref.~\cite{4430belle2} the Belle Collaboration updated their
predictions for the mass and width of the $Z(4430)$ resonance:
$M\,=\,4443^{+15}_{-12}{\rm \thinspace (stat)}^{+19}_{-13}{\rm\thinspace (syst)}$ MeV and 
$\Gamma\,=\,107^{+86}_{-43}{\rm \thinspace (stat)}^{+74}_{-56}{\rm \thinspace (syst)}$ MeV.

The {\it BABAR} Collaboration studied the decays
$\bar B^{-,0}\, \rightarrow\,K^{0,+}\pi^-\psi(2s)$
and $\bar B^{-,0}\, \rightarrow\,K^{0,+}\pi^-J/\psi$,
but they did not see a $Z(4430)^-$
signal~\cite{4430babar}. They derived upper limits for branching fractions 
that are yet compatible with the mentioned results of the Belle Collaboration.
In Ref.~\cite{4430belle3} the Belle Collaboration reported on 
the spin and parity of the $Z(4430)^-$ state
constrained from a full amplitude analysis of the
$B^0\rightarrow\psi(2s) K^+\pi^-$ decay with
$\psi(2s)\rightarrow\mu^+\mu^-$ or
$e^+e^-$.
They found that the $Z(4430)^-$ being a $J^P\,=\,1^+$-state was favored
over the next likely state ($0^-$) with a significance of 3.4$\sigma$.

Furthermore, the Belle Collaboration did estimate for the
product of branching fractions:
$\mathcal{B}(\bar B^0 \rightarrow K^-Z(4430)^+)\times                                  
\mathcal{B}(Z(4430)^+\rightarrow\pi^+\psi(2s))\,=\,                                       
(3.2^{+1.8}_{-0.9}{\rm \thinspace (stat)}^{+5.3}_{-1.6}
{\rm \thinspace (syst)})\times 10^{-5}$.
The {\it BABAR} Collaboration studied the decays
$\bar B^{-,0}\, \rightarrow\,K^{0,+}\pi^-\psi(2s)$
and $\bar B^{-,0}\, \rightarrow\,K^{0,+}\pi^-J/\psi$,
but they did not see a $Z(4430)^-$
signal~\cite{4430babar}. They derived upper limits for branching fractions 
that are yet compatible with the mentioned results of the Belle Collaboration.

In Ref.~\cite{4430belle3} the Belle Collaboration reported on 
the spin and parity of the $Z(4430)^-$ state
constrained from a full amplitude analysis of the
$B^0\rightarrow\psi(2s) K^+\pi^-$ decay with
$\psi(2s)\rightarrow\mu^+\mu^-$ or
$e^+e^-$.
They found that the $Z(4430)^-$ being a $J^P\,=\,1^+$-state was favored
over the next likely state ($0^-$) with a significance of 3.4$\sigma$.

In Ref.~\cite{4430lhcb1} the LHCb Collaboration confirmed
the $Z(4430)^-$ signal in the $\psi(2s)\pi^-$-spectrum of the decay
$B^0\rightarrow\psi^\prime K^+\pi^-$, and determined unambiguously
the spin parity $J^P\,=\,1^+$ of the $Z(4430)$ state~\cite{4430lhcb1}.
They determined the following values for the mass and width of
the $Z(4430)^-$ state:
$M\,=\,4475 \pm 7{\rm \thinspace (stat)}^{+15}_{-25}{\rm \thinspace (syst)}$~MeV and
$\Gamma\,=\,172 \pm 13{\rm \thinspace (stat)}^{+37}_{-34}      
{\rm \thinspace (syst)}$~MeV~\cite{4430lhcb1}.
In Ref.~\cite{4430lhcb2} the LCHb Collaboration concluded
that the only possible explanation for internal structure of
the $Z(4430)$ state is a four-quark $ccud$ bound state.

In Ref.~\cite{4430belle4} the Belle Collaboration reported, that they also found
a $Z^+(4430)$ signal in the $J/\psi \pi^+$spectrum of the decay
$\bar B^0\,\rightarrow\,J/\psi K^-\pi^+$. They report product branching fraction
$\mathcal{B}(\bar B^0 \rightarrow K^-Z(4430)^+)\times                                     
\mathcal{B}(Z(4430)^+\rightarrow\pi^+J/\psi)\,=\,                                        
(5.4^{+4.0}_{-1.0}{\rm \thinspace (stat)}^{+1.1}_{-0.9}{\rm \thinspace (syst)})
\times 10^{-5}$.
If one compares this value with the corresponding product branching fraction
for the $\psi(2s)$ particle (see above) and assumes that the decay rates are invariant under
charge conjugation, one can derive an estimation for the branching ratio of the two decay channels
of the $Z(4430)^\pm$.
We do not know how the errors of the values correlate,
so we only do a rough estimation of the errors by dividing
the upper limit of the one product branching fraction by the
lower limit of the other one and vice versa, taking into account statistical
and systematical error both in one step.
We get 
\bea
R_Z = \frac{\Gamma(Z(4430)^\pm\rightarrow\pi^\pm\psi(2s))}                                      
{\Gamma(Z(4430)^\pm\rightarrow\pi^\pm J/\psi)}\,\simeq\, 11.1^{+18}_{-8.6} \,. 
\ena 

In the present paper we critically check the tetraquark picture for 
both the $Z_c(3900)$ and $X(5568)$ states by analyzing their strong decays. 
In our consideration we use the covariant quark model
proposed in \cite{Branz:2009cd} and used 
in Refs.~\cite{Dubnicka:2010kz,Dubnicka:2011mm}
to describe the properties of the $X(3872)$ state as a tetraquark state.
First, we employ an interpretation of the $Z_c(3900)$ state 
as the isospin 1 partner of the $X(3872)$ as was suggested in 
Refs.~\cite{Dias:2013xfa} and~\cite{Faccini:2013lda}.
We calculate the partial widths of the decays
$Z_c^+(3900)\to\Jpsi\pi^+$, $\eta_c\rho^+$, and $\bar D^0D^{\ast\,+}$, 
$\bar D^{\ast\,0}D^{+}$. We find that for a relatively small model
size parameter $\Lambda_{Z_c} \sim 1.4$~GeV one can reproduce 
the central values 
for the partial widths of the decays $Z^+_c\to\Jpsi\pi^+$, $\eta_c\rho^+$
as they were also obtained 
in Refs.~\cite{Dias:2013xfa,Faccini:2013lda}. 
It turns out that, in our model, the leading Lorentz metric structure in the
matrix elements
describing the decays $Z_c(3900)\to\bar DD^{\ast}$ vanishes analytically.
This results in a significant suppression of these decay widths by
the smallness of the relevant phase space factor ${\bf|q|}^5$. 
Since the experimental data \cite{Ablikim:2013xfr} show
that the $Z_c(3900)$ has a much more stronger coupling to $DD^\ast$ than
to $\Jpsi\pi$, one has to conclude that the tetraquark-type current  
for the $Z_c(3900)$ is in discord with experiment.
As an alternative we employ a molecular-type four-quark current to describe 
the decays of the $Z_c(3900)$ state. In this case we find that for a
relatively large size parameter $\Lambda_{Z_c} \sim 3.3$~GeV 
one can obtain the partial widths of the decays $Z_c^+(3900)\to\bar DD^{\ast}$ 
at the order  $\sim 15$~MeV for each mode. At the same time
the partial widths for decays $Z_c^+(3900)\to\Jpsi\pi^+\,, \eta_c\rho^+$
are suppressed by a factor of $6-7$ in accordance with experimental
data~\cite{Ablikim:2013xfr}.  

Let us stress, that in our manuscript we consider exotic mesons 
in the four-quark picture with the use of two possible configuration 
of quarks in these states.
Note that molecular configuration does not mean that a specific size of
the state with such structure is more compact than the tetraquark
configuration. In this sense it is differed from hadronic molecules -
extended object with clear separation of two hadrons - the constituents 
building the exotic state. Such hadronic molecules (extended objects)
with smaller size parameter (of order of 1-2 GeV) have been considered
some of us in the phenomenological Lagrangian approach based on the
composite structure of exotic states as bound states of separate 
hadrons~\cite{hmappr}. 
In the present manuscript, for the first time in order
to distinguish both configurations we vary the size parameter
in the same region 3.2 - 3.4 GeV, which is guided by experimental data. 
Also we would like to mention that the size parameter is not directly
related to the size of a hadron like e.g., in potential approaches.
Indeed, our size parameter is related to the physical quantities
like electromagnetic radii, slope of the form factors, etc. 

Then, we test the tetraquark picture for the 
$X(5568)$ structure by analyzing its strong one-pion decay.
We found that for a mass of $5568$~MeV one can fit the experimental 
decay width by using the value of size parameter
$\Lambda_{X_b}\sim 1.4$~GeV. In the case of a larger mass of  
$5771$~MeV~\cite{Ali:2016gdg} one finds  $\Lambda_{X_b}\sim 1.7$~GeV.
Finally, we consider the decays of the $Z(4430)$ state 
$Z(4430)^+ \to J/\psi + \pi^+$, $Z(4430)^+ \to \psi(2s) + \pi^+$, 
and $Z(4430)^+ \to D^{\ast\,+} + \bar D^{\ast\,0}$  
in the tetraquark picture. 

The paper is organized as follows. In Sec.~II, we consider
the $Z_c(3900)$ state, a four-quark state, as a compact
tetraquark bounded by color diquark and antidiquark.
In Sec.~III we test a molecular-type four-quark structure of
the $Z_c(3900)$ state. In Sec.~IV we present study of the $X(5568)$
exotic state as the tetraquark four-quark state.
In Sec.~V we apply the tetraquark model for the $Z(4430)$ state.
Finally, in Sec.~VI we summarize and conclude our results.


\section{The $Z_c(3900)$ as a four-quark state with a tetraquark-type
  current}

Let us first interpret $Z_c(3900)$ as the isospin 1 partner
of the $X(3872)$ as was suggested in Refs.~\cite{Dias:2013xfa} and 
\cite{Faccini:2013lda}. Then the quantum numbers for the neutral state
are  $I^G(J^{PC}) = 1^+(1^{+-})$. Accordingly the interpolating current for
the $Z^+_c(3900)$ state is given by 
\be
J^\mu = \frac{i}{\sqrt{2}}\varepsilon_{abc}\varepsilon_{dec}
\left[  (u^T_a C\gamma_5 c_b)(\bar d_d\gamma^\mu C \bar c^T_e)
      - (u^T_a C\gamma^\mu c_b)(\bar d_d\gamma_5 C \bar c^T_e)\right]
\label{eq:Zc-tet-cur}
\en 
We employ a charge conjugation matrix in the form of
$C=\gamma^0\gamma^2$, i.e., without a factor  ``i'' as is usually employed.
This allows one to simplify the calculations because of
$C=C^\dagger=C^{-1}=-C^T$, $C\Gamma^TC^{-1}=\pm\Gamma$ ($"+"$ for
$\Gamma=S,P,A$ and $"-"$ for $\Gamma=V,T$). In what follows we
drop the superscript ``T'' (transpose) from the spinors
to avoid a complication of notation.  

The nonlocal version of the four-quark interpolating current
reads
\bea
J^\mu_{Z_c}(x) &=& \int\! dx_1\ldots \int\! dx_4 
\delta\left(x-\sum\limits_{i=1}^4 w_i x_i\right) 
\Phi_{Z_c}\Big(\sum\limits_{i<j} (x_i-x_j)^2 \Big)
J^\mu_{4q}(x_1,\ldots,x_4),
\label{eq:nonlocal-cur}\\
J^\mu_{4q}&=&
\frac{i}{\sqrt{2}}\, \varepsilon_{abc}\varepsilon_{dec} \,
\Big\{\, [u_a(x_4)C\gamma_5 c_b(x_1)][\bar d_d(x_3)\gamma^\mu C \bar c_e(x_2)]
\nn
&&
\phantom{\frac{i}{\sqrt{2}}\,\varepsilon_{abc}\varepsilon_{dec}}  
-  [u_a(x_4)C\gamma^\mu c_b(x_1)][\bar q_d(x_3)\gamma_5 C \bar c_e(x_2)]
\,\Big\}
\nonumber
\ena 
where $w_i=m_i/\sum_{j=1}^4 m_j$.
The numbering of the coordinates
$x_i$ is chosen such that one has a convenient arrangement of
vertices and propagators in the Feynman diagrams to be calculated.
The effective interaction Lagrangian describing the coupling
of the meson $Z_c$ to its constituent quarks is written in the form
\be
{\cal L}_{\rm int} = g_{Z_c}\,Z_{c,\,\mu}(x)\cdot J^\mu_{Z_c}(x) + \text{H.c.}
\label{eq:lag}
\en     

The Fourier transform of the vertex  function
$\Phi_{Z_c}\Bigl(\sum\limits_{i<j} ( x_i - x_j )^2 \Bigr)$ can be calculated
by using appropriately chosen Jacobi coordinates 
\be
x_i = x + \sum\limits_{j=1}^3 w_{ij} \rho_j
\label{eq:Jacobi}
\en
where
\[
\begin{array}{lll}
w_{11} = +\,\tfrac{2w_2+w_3+w_4}{2\sqrt{2}} \qquad &
w_{12} = -\,\tfrac{w_3- w_4}{2\sqrt{2}}   \qquad  &
w_{13} = +\,\tfrac{w_3+w_4}{2}               
\\
w_{21} = -\,\tfrac{2w_1+w_3+w_4}{2\sqrt{2}}  &
w_{22} = -\,\tfrac{w_3- w_4}{2\sqrt{2}}      &
w_{23} = +\,\tfrac{w_3+w_4}{2}               
\\
w_{31} = -\,\tfrac{w_1-w_2}{2\sqrt{2}}       &
w_{32} =  +\,\tfrac{w_1+w_2+2w_4}{2\sqrt{2}}  &
w_{33} =  -\,\tfrac{w_1+w_2}{2}              
\\
w_{41} = -\,\tfrac{w_1-w_2}{2\sqrt{2}}       &
w_{42} = -\,\tfrac{w_1+w_2+2w_3}{2\sqrt{2}}   &
w_{43} = -\,\tfrac{w_1+w_2}{2}              
\end{array}
\]
It is straightforward to check that 
$x=\sum\limits_{i=1}^4 x_i w_i$, and 
$\sum\limits_{1\le i< j\le 4} (x_i-x_j)^2 =\sum\limits_{i=1}^3 \rho_i^2.$ 
The vertex function is then written as
\bea
\Phi_{Z_c}\Bigl(\sum\limits_{i<j} ( x_i - x_j )^2 \Bigr)
&=&
\int\frac{d\vec\omega}{(2\pi)^{12}}
e^{-i\,\vec\rho\,\vec\omega}\, \widetilde\Phi_{Z_c}(-\vec\omega^{\,2})\,, 
\ena
where the vertex function in momentum space is chosen to have
a Gaussian form
\be
\widetilde\Phi_{Z_c}(-\vec\omega^{\,2}) = \exp(\vec\omega^{\,2}/\Lambda^2_{Z_c} )
\en
with the $\Lambda^2_{Z_c}$ being an adjustable size parameter.

The coupling constant $g_{Z_c}$ in Eq.~(\ref{eq:lag}) is determined by
the normalization condition called {\it the compositeness condition} 
(see Refs.~\cite{Efimov:1993zg} and \cite{Branz:2009cd} for details), 
\be
\label{eq:Z=0}
Z_{Z_c} = 1-g^2_{Z_c}\,\widetilde\Pi_{Z_c}^\prime(m^2_{Z_c})=0,
\en
where $\Pi_{Z_c}(p^2)$ is the scalar part of the vector-meson mass operator
\bea
\widetilde\Pi^{\mu\nu}_{Z_c}(p) &=& g^{\mu\nu} \widetilde\Pi_{Z_c}(p^2) 
                            + p^\mu p^\nu \widetilde\Pi^{(1)}_{Z_c}(p^2),
\nn
\widetilde\Pi_{Z_c}(p^2) &=& 
\frac13\left(g_{\mu\nu}-\frac{p_\mu p_\nu}{p^2}\right)\Pi^{\mu\nu}_{Z_c}(p).
\label{eq:mass}
\ena
The Fourier transform of the $Z_c$-tetraquark mass operator reads
\bea
\Pi_{Z_c}^{\mu\nu}(p) &=& 
6\,\prod\limits_{i=1}^3\int\!\!\frac{d^4k_i}{(2\pi)^4i}\,
\widetilde\Phi_{Z_c}^2\left(-\,\vec\omega^{\,2}\right) 
\nn
&\times& \Big\{
\,\,\Tr\left[S_4(\hat k_4)\gamma_5 S_1(\hat k_1) \gamma_5 \right]
\Tr\left[S_3(\hat k_3)\gamma^\mu S_2(\hat k_2)\gamma^\nu \right] 
\nn
&&
+\,\Tr\left[S_4(\hat k_4)\gamma^\nu S_2(\hat k_2) \gamma^\mu \right]
\Tr\left[S_3(\hat k_3)\gamma_5 S_1^(\hat k_1)\gamma_5 \right] 
\Big\}
\label{eq:Fourier-mass}
\ena
where $\hat k_1=k_1-w_1 p$, $\hat k_2=k_2-w_2 p$,  $\hat k_3=k_3+w_3 p$,
      $\hat k_4=k_1+k_2-k_3+w_4 p$, and
$\vec\omega^{\,2}=1/2\,(k_1^2+k_2^2+k_3^2+k_1k_2-k_1k_3-k_2k_3)$.    
Details of the calculation can be found in our previous papers,
e.g.~\cite{Dubnicka:2010kz,Dubnicka:2011mm}.
 
The matrix elements of the decays 
 $Z_c^+\to \Jpsi+\pi^+$ and $Z_c^+\to \eta_c+\rho^+$ 
are given by
\bea\label{Zc_Jpsi_pi1}
&&
M^{\mu\nu}
\left( Z_c(p,\epsilon^\mu_p) \to \Jpsi(q_1,\epsilon^\nu_{q_1})+\pi^+(q_2)\right)
= \frac{6}{\sqrt{2}}\,g_{Z_c}g_{\Jpsi}g_{\pi}
\nn
&\times&
\int\!\!\frac{d^4k_1}{(2\pi)^4i}\,\int\!\!\frac{d^4k_2}{(2\pi)^4i}\,
\widetilde\Phi_{Z_c}\left(-\,\vec\eta^{\,2}\right)
\widetilde\Phi_{\Jpsi}\left(-\,(k_1+v_2 q_1)^2\right)
\widetilde\Phi_{\pi}\left(-\,(k_2+u_4 q_2)^2\right)
\nn
&\times& \Big\{
\,\,\Tr\left[ \gamma_5 S_4(k_2)\gamma_5 S_3(k_2+q_2) 
              \gamma^\mu S_2(k_1)\gamma^\nu S_1(k_1+q_1) \right] 
\nn
&&
+\,\Tr\left[ \gamma^\mu S_4(k_2)\gamma_5 S_3(k_2+q_2) 
             \gamma_5 S_2(k_1)\gamma^\nu S_1(k_1+q_1) \right] 
\Big\}
\nn[2ex]
&=& A_{\Jpsi\pi}\,g^{\mu\nu} + B_{\Jpsi\pi}\,q_1^\mu q_2^\nu\,,
\label{eq:decay-CC1}
\ena

\bea
&&
M^{\mu\alpha}
\left( Z_c(p,\epsilon^\mu_p) \to \eta_c(q_1)+\rho(q_2,\epsilon^\alpha_{q_2})\right)
=  \frac{6}{\sqrt{2}}\,g_{Z_c}g_{\eta_c}g_{\rho}
\nn
&\times&
\int\!\!\frac{d^4k_1}{(2\pi)^4i}\,\int\!\!\frac{d^4k_2}{(2\pi)^4i}\,
\widetilde\Phi_{Z_c}\left(-\,\vec\eta^{\,2}\right)
\widetilde\Phi_{\eta_c}\left(-\,(k_1+v_2 q_1)^2\right)
\widetilde\Phi_{\rho}\left(-\,(k_2+u_4 q_2)^2\right)
\nn
&\times& \Big\{
\,\,\Tr\left[ \gamma_5 S_4(k_2)\gamma^\alpha S_3(k_2+q_2) 
              \gamma^\mu S_2(k_1)\gamma_5 S_1(k_1+q_1) \right] 
\nn
&&
+\,\Tr\left[  \gamma^\mu S_4(k_2)\gamma^\alpha S_3(k_2+q_2) 
               \gamma_5 S_2(k_1)\gamma_5 S_1(k_1+q_1) \right] 
\Big\}
\nn[2ex]
&=& A_{\eta_c\rho}\,g^{\mu\alpha} - B_{\eta_c\rho}\,q_2^\mu q_1^\alpha\,.
\label{eq:decay-CC2}
\ena
The argument of the $Z_c$-vertex function is given by
\bea
\vec\eta^{\,2} &=& \eta_1^2+\eta_2^2+\eta_3^2,
\nn
\eta_1 &=& + \frac{1}{2\sqrt{2}}\left(2k_1+(1-w_1+w_2) q_1 -(w_1-w_2) q_2 \right),
\nn
\eta_2 &=& + \frac{1}{2\sqrt{2}}\left(2k_2-(w_3-w_4) q_1 +(1-w_3+w_4) q_2 \right),
\nn
\eta_3 &=& + \frac{1}{2}\left((w_3+w_4) q_1 - (w_1+w_2) q_2 \right).
\label{eq:Zc-arg-CC}
\ena
The quark masses are specified as $m_1=m_2=m_c$,  $m_3=m_4=m_d=m_u$,
and the two-body reduced masses as $v_1=m_1/(m_1+m_2)$,  $v_2=m_2/(m_1+m_2)$, 
$u_3=m_3/(m_3+m_4)$, and $u_4=m_4/(m_3+m_4)$. 
 
The matrix elements of the decays 
 $Z_c^+\to \bar D^0+D^{\ast\,+}$ and $Z_c^+\to \bar D^{\ast\, 0}+D^{+}$
read
\bea
&&
M^{\mu\nu}
\left( Z_c(p,\epsilon^\mu_p)
\to\bar D^0(q_1)+D^{\ast\,+}(q_2,\epsilon^\nu_{q_2})\right)
= \frac{6}{\sqrt{2}}\,g_{Z_c}g_{D}g_{D^\ast}
\nn
&\times&
\int\!\!\frac{d^4k_1}{(2\pi)^4i}\,\int\!\!\frac{d^4k_2}{(2\pi)^4i}\,
\widetilde\Phi_{Z_c}\left(-\,\vec\delta^{\,2}\right)
   \widetilde\Phi_{D}\left(-\,(k_2+v_2 q_2)^2\right)
\widetilde\Phi_{D^\ast}\left(-\,(k_1+u_1 q_2)^2\right)
\nn
&\times& \Big\{
\,\,\Tr\left[ \gamma_5 S_4(k_2+q_1)\gamma_5 S_1(k_1) 
              \gamma^\nu S_3(k_1+q_2)\gamma^\mu S_2(k_2)  \right] 
\nn
&&
-\,\Tr\left[ \gamma_5 S_4(k_2+q_1)\gamma^\mu S_1(k_1) 
              \gamma^\nu S_3(k_1+q_2)\gamma_5 S_2(k_2)  \right] 
\Big\}
\nn[2ex]
&=& A_{\bar DD^\ast}\,g^{\mu\nu} - B_{\bar DD^\ast}\,q_2^\mu q_1^\nu\,,
\label{eq:decay-DD1}
\ena
\bea
&&
M^{\mu\alpha}
\left( Z_c(p,\epsilon^\mu_p) 
\to\bar D^{\ast\,0} (q_1,\epsilon^\alpha_{q_1})+D^+(q_2,)\right)
=\frac{6}{\sqrt{2}}\,g_{Z_c}g_{D^\ast}g_{D}
\nn
&\times&
\int\!\!\frac{d^4k_1}{(2\pi)^4i}\,\int\!\!\frac{d^4k_2}{(2\pi)^4i}\,
\widetilde\Phi_{Z_c}\left(-\,\vec\delta^{\,2}\right)
\widetilde\Phi_{D^\ast}\left(-\,(k_1+\hat v_1 q_1)^2\right)
\widetilde\Phi_{D}\left(-\,(k_2+\hat u_4 q_2)^2\right)
\nn
&\times& \Big\{
\,\,\Tr\left[ S_4(k_2+q_1)\gamma_5 S_1(k_1) 
              \gamma_5 S_3(k_1+q_2)\gamma^\mu S_2(k_2)\gamma^\alpha \right] 
\nn
&&
-\,\Tr\left[ S_4(k_2+q_1)\gamma^\mu S_1(k_1) 
              \gamma_5 S_3(k_1+q_2)\gamma_5 S_2(k_2)\gamma^\alpha \right] 
\Big\}
\nn[2ex]
&=& A_{D^\ast D}\,g^{\mu\alpha} + B_{D^\ast D}\,q_1^\mu q_2^\alpha\,.
\label{eq:decay-DD2}
\ena
The argument of the  $Z_c$-vertex function is given by
\bea
\vec\delta^{\,2} &=& \delta_1^2+\delta_2^2+\delta_3^2,
\nn
\delta_1 &=& - \frac{1}{2\sqrt{2}}
\left(k_1-k_2+(w_1-w_2) (q_1+q_2) \right),
\nn
\delta_2 &=& + \frac{1}{2\sqrt{2}}
\left(k_1-k_2-(1+w_3-w_4) q_1 +(1-w_3+w_4) q_2 \right),
\nn
\delta_3 &=& - \frac{1}{2}\left(k_1+k_2+(w_1+w_2) (q_1+q_2) \right).
\label{eq:arg-DD2}
\ena
The quark masses are specified as $m_1=m_2=m_c$,  $m_3=m_4=m_d=m_u$,
and the two-body reduced masses as
$\hat v_2=m_2/(m_2+m_4)$,  $\hat v_4=m_4/(m_2+m_4)$, 
$\hat u_1=m_1/(m_1+m_3)$, and $\hat u_3=m_3/(m_1+m_3)$. 

We finally calculate the two-body decay widths. The relevant
spin kinematical formulas have been collected in the appendix.
Note that momentum of the daughter vector particle is chosen to
be $q_1$ in Eq.~(\ref{eq:A1}). In addition the matrix element is expressed
through the dimensionless invariant amplitudes $A_1$, and $A_2$ in
Eq.~(\ref{eq:A3}). In order to adjust the notation in Eqs.~(\ref{eq:decay-CC1}),
(\ref{eq:decay-CC2}), (\ref{eq:decay-DD1}), and (\ref{eq:decay-DD2})
to those given in the appendix, one has to replace $q_1\leftrightarrow q_2$
in Eqs.~(\ref{eq:decay-CC2}) and (\ref{eq:decay-DD1}) and then 
introduce the dimensionless  form factors
$A_1=A/m$ and $A_2=\pm m\, B$\,\,\, ($p^2=m^2$) where the sign $``+''$ stands 
for Eqs.~(\ref{eq:decay-CC1}) and (\ref{eq:decay-DD2})  
and $``-''$ for Eq.~(\ref{eq:decay-CC2}) and (\ref{eq:decay-DD1}), respectively.
The expressions for helicity amplitudes via $A_1$ and $A_2$
are given in Eq.~(\ref{eq:A5}). 
The two-body decay widths are now calculated using Eq.~(\ref{eq:A9}). 

As a consequence of the subtraction of the two traces in the matrix elements
in Eqs.~(\ref{eq:decay-DD1}) and (\ref{eq:decay-DD2}) we found that
$A_{DD^\ast}=A_{D^\ast D}\equiv 0$ analytically. This results in 
a significant suppression of the decay widths due to 
the $D$--wave suppression factor of ${\bf|q_1|}^5$. 
In the calculation of the quark-loop diagrams 
we have only one free parameter $\Lambda_{Z_c}$, the size parameter of 
the $Z_c$ state. 
The other model parameters have been fixed in previous 
papers~\cite{Branz:2009cd}-\cite{Dubnicka:2011mm},\cite{Ivanov:2006ni} 
from analysis of hadron processes involving light and heavy quarks,  
\be
\def\arraystretch{2}
\begin{array}{ccccccc}
     m_{u/d}        &      m_s        &      m_c       &     m_b & \lambda  &
\\ \hline
 \  0.241 \    &  \  0.428 \   &  \ 1.67 \   &  \ 5.05 \   &
\  0.181 \   &  \ {\rm GeV.}
\end{array}
\label{eq: fitmas}
\en
Here  $m_q$ are the constituent quark masses and $\lambda$ is 
an infrared cutoff parameter responsible for the quark confinement.  
The size parameters of the $\pi$ $\rho$, $D$, $D^\ast$, $J/\psi$, and 
$\eta_c$ have been fixed as 
\be
\def\arraystretch{2}
\begin{array}{ccccccc}
   \Lambda_\pi  &   \Lambda_\rho &   \Lambda_D  &   
   \Lambda_{D^\ast}  &   \Lambda_{J/\psi} &   \Lambda_{\eta_c}  &   
\\ \hline
 \  0.871 \    &  \  0.624 \   &  \ 1.600 \   &
 \  1.529 \    &  \  1.738 \   &  \ 3.777 \   & \text{GeV.}
\end{array}
\label{eq: fitsize}
\en

For the $Z_c(3900)$ mass we use the actual value 3.886 GeV.
We adjust the size parameter $\Lambda_{Z_c}$
in such a way as to be close to the central value for the decay
$Z^+_c\to\Jpsi+\pi^+$ obtained in Refs.~\cite{Dias:2013xfa,Faccini:2013lda}.
If the parameter $\Lambda_{Z_c}$ is varied in the region
$\Lambda_{Z_c}=2.25\pm 0.10$~GeV the numerical values of the decay widths
vary as
\bea
\Gamma(Z^+_c\to\Jpsi+\pi^+) &=& (27.9^{+6.3}_{-5.0})\,\text{MeV}\,,
\nn
\Gamma(Z^+_c\to\eta_c+\rho^+) &=& (35.7^{+6.3}_{-5.2})\,\text{MeV}\,,
\nn
\Gamma(Z^+_c\to \bar D^0 + D^{\ast\, +}) &\propto & 10^{-8}\,\text{MeV}\,,
\nn
\Gamma(Z^+_c\to \bar D^{\ast\, 0} + D^+) &\propto & 10^{-8}\,\text{MeV}\,.
\label{eq:Zc-tetra-values}
\ena
Here and in the following an increasing of the size parameter leads
to a decreasing of the decay width.
Since the experimental data~\cite{Ablikim:2013xfr} show
that the $Z_c(3900)$ has a much more stronger coupling to $DD^\ast$ than
$\Jpsi\pi$, one has to conclude that the tetraquark-type current
for $Z_c(3900)$ is in discord with experiment. 

Moreover, we expect that a realistic value of the size parameter $\Lambda_{Z_c}$ is 
about 3 GeV. Using $\Lambda_{Z_c} = 3.3 \pm 1.1$ GeV we get a significant 
suppression for the $Z^+_c\to\Jpsi+\pi^+$ and $Z^+_c\to\eta_c+\rho^+$ modes, 
and the rates for the modes 
$Z^+_c\to \bar D^0 + D^{\ast\, +}$ and $Z^+_c\to \bar D^{\ast\, 0} + D^+$ 
become much more negligible 
\bea
\Gamma(Z^+_c\to\Jpsi+\pi^+) &=& (4.3^{+0.7}_{-0.6})\,\text{MeV}\,,
\nn
\Gamma(Z^+_c\to\eta_c+\rho^+) &=& (8.0^{+1.2}_{-1.0})\,\text{MeV}\,,
\nn
\Gamma(Z^+_c\to \bar D^0 + D^{\ast\, +}) &\propto & 10^{-9}\,\text{MeV}\,,
\nn
\Gamma(Z^+_c\to \bar D^{\ast\, 0} + D^+) &\propto & 10^{-9}\,\text{MeV}\,.
\label{eq:Zc-tetra-values2}
\ena

\section{The $Z_c(3900)$ as a four-quark state with a molecular-type current}

We describe the $Z^+_c(3900)$ as the charged particle
in the isotriplet with a molecular-type current given by 
(see Ref.~\cite{Nielsen:2009uh})
\be
J^\mu = \frac{1}{\sqrt{2}} 
\left[ (\bar d \gamma_5 c) (\bar c\gamma^\mu u)
      +(\bar d \gamma^\mu c)(\bar c\gamma_5  u) \right]
\label{eq:mol-cur}.
\en 
Its  nonlocal generalization is given by
\bea
J^\mu_{Z_c}(x) &=& \int\! dx_1\ldots \int\! dx_4 
\delta\left(x-\sum\limits_{i=1}^4 w_i x_i\right) 
\Phi_{Z_c}\Big(\sum\limits_{i<j} (x_i-x_j)^2 \Big)
J^\mu_{4q}(x_1,\ldots,x_4),
\label{eq:nonlocal-mol-cur}\\
J^\mu_{4q}&=&  \frac{1}{\sqrt{2}} 
\Big\{
 (\bar d(x_3) \gamma_5   c(x_1)) (\bar c(x_2) \gamma^\mu u(x_4) )
+(\bar d(x_3) \gamma^\mu c(x_1))  (\bar c(x_2) \gamma_5  u(x_4) )
\,\Big\}
\nonumber
\ena 

The Fourier transform of the $Z_c$ mass operator is written as
\bea
\Pi_{Z_c}^{\mu\nu}(p) &=& 
\frac{9}{2} \,\prod\limits_{i=1}^3\int\!\!\frac{d^4k_i}{(2\pi)^4i}\,
\widetilde\Phi_{Z_c}^2\left(-\,\vec\omega^{\,2}\right) 
\nn
&\times& \Big\{
\,\,\Tr\left[\gamma_5  S_1(\hat k_1)\gamma_5 S_3(\hat k_3)  \right]
    \Tr\left[\gamma^\mu S_4(\hat k_4)\gamma^\nu S_2(\hat k_2)\right] 
\nn
&&
+\, \Tr\left[\gamma^\mu S_1(\hat k_1)\gamma^\nu S_3(\hat k_3)  \right]
    \Tr\left[\gamma_5 S_4(\hat k_4)\gamma_5 S_2(\hat k_2)\right] 
\Big\}
\label{eq:Fourier-mol-mass}
\ena
with  $\hat k_i$ and $\vec\omega^{\,2}$ being defined as in the previous
section.
 
The matrix elements of the decays  $Z^+_c\to \Jpsi+\pi^+$ and
 $Z^+_c\to\eta_c+\rho^+$ are given by
\bea
&&
M^{\mu\nu}
\left( Z_c(p,\epsilon^\mu_p) \to \Jpsi(q_1,\epsilon^\nu_{q_1})+\pi^+(q_2)\right)
= \frac{3}{\sqrt{2}}\,g_{Z_c}g_{\Jpsi}g_{\pi}
\nn
&\times&
\int\!\!\frac{d^4k_1}{(2\pi)^4i}\,\int\!\!\frac{d^4k_2}{(2\pi)^4i}\,
\widetilde\Phi_{Z_c}\left(-\,\vec\eta^{\,2}\right)
\widetilde\Phi_{\Jpsi}\left(-\,(k_1+v_1 q_1)^2\right)
\widetilde\Phi_{\pi}\left(-\,(k_2+u_4 q_2)^2\right)
\nn
&\times& \Big\{
\,\,\Tr\left[ \gamma_5   S_1(k_1)\gamma^\nu S_2(k_1+q_1) 
              \gamma^\mu S_4(k_2)\gamma_5 S_3(k_2+q_2) \right] 
\nn
&&
+\,\Tr\left[ \gamma^\mu   S_1(k_1)\gamma^\nu S_2(k_1+q_1) 
              \gamma_5 S_4(k_2)\gamma_5 S_3(k_2+q_2) \right] 
\Big\}
\nn[2ex]
&=& A_{\Jpsi\pi}\,g^{\mu\nu} + B_{\Jpsi\pi}\,q_1^\mu q_2^\nu\,,
\label{eq:decay-mol-CC1}
\ena
\bea
&&
M^{\mu\alpha}
\left( Z_c(p,\epsilon^\mu_p) \to \eta_c(q_1)+\rho(q_2,\epsilon^\alpha_{q_2})\right)
= \frac{3}{\sqrt{2}}\,g_{Z_c}g_{\eta_c}g_{\rho}
\nn
&\times&
\int\!\!\frac{d^4k_1}{(2\pi)^4i}\,\int\!\!\frac{d^4k_2}{(2\pi)^4i}\,
\widetilde\Phi_{Z_c}\left(-\,\vec\eta^{\,2}\right)
\widetilde\Phi_{\eta_c}\left(-\,(k_1+v_1 q_1)^2\right)
\widetilde\Phi_{\rho}\left(-\,(k_2+u_4 q_2)^2\right)
\nn
&\times& \Big\{
\,\,\Tr\left[ \gamma_5 S_1(k_1)\gamma_5 S_2(k_1+q_1) 
              \gamma^\mu S_4(k_2)\gamma^\alpha S_3(k_2+q_2) \right] 
\nn
&&
+\,\Tr\left[  \gamma^\mu S_1(k_1)\gamma_5 S_2(k_1+q_1) 
              \gamma_5 S_4(k_2)\gamma^\alpha S_3(k_2+q_2) \right] 
\Big\}
\nn[2ex]
&=& A_{\eta_c\rho}\,g^{\mu\alpha} - B_{\eta_c\rho}\,q_2^\mu q_1^\alpha\,.
\label{eq:decay-mol-CC2}
\ena
The argument of the $Z_c$-vertex function $\vec\eta^{\,2}$
and the specification of the quark masses are identical to those given in
the previous section.

The matrix elements of the decays $Z^+_c\to \bar D^{0}+D^{\ast\,+}$ and
 $Z^+_c\to \bar D^{\ast\, 0}+D^{+}$ read
\bea
&&
M^{\mu\nu}
\left( Z_c(p,\epsilon^\mu_p)
\to\bar D^0(q_1)+D^{\ast\,+}(q_2,\epsilon^\nu_{q_2})\right)
=\frac{9}{\sqrt{2}}\,g_{Z_c}g_{D}g_{D^\ast}
\nn
&\times&
\int\!\!\frac{d^4k_1}{(2\pi)^4i}\,\int\!\!\frac{d^4k_2}{(2\pi)^4i}\,
\widetilde\Phi_{Z_c}\left(-\,\vec\delta^{\,2}\right)
   \widetilde\Phi_{D}\left(-\,(k_2+v_4 q_1)^2\right)
\widetilde\Phi_{D^\ast}\left(-\,(k_1+u_1 q_2)^2\right)
\nn
&\times& \Big\{
\,\,\Tr\left[ \gamma^\mu S_1(k_1)\gamma^\nu S_3(k_1+q_2)  \right]  
    \Tr\left[ \gamma_5 S_4(k_2)\gamma_5 S_2(k_2+q_1)  \right] 
\Big\}
\nn[2ex]
&=& A_{\bar DD^\ast}\,g^{\mu\nu} - B_{\bar DD^\ast}\,q_2^\mu q_1^\nu\,,
\label{eq:decay-mol-DD1}
\ena
\bea
&&
M^{\mu\alpha}
\left( Z_c(p,\epsilon^\mu_p) 
\to\bar D^{\ast\,0} (q_1,\epsilon^\alpha_{q_1})+D^+(q_2,)\right)
=\frac{9}{\sqrt{2}}\,g_{Z_c}g_{D^\ast}g_{D}
\nn
&\times&
\int\!\!\frac{d^4k_1}{(2\pi)^4i}\,\int\!\!\frac{d^4k_2}{(2\pi)^4i}\,
\widetilde\Phi_{Z_c}\left(-\,\vec\delta^{\,2}\right)
\widetilde\Phi_{D^\ast}\left(-\,(k_1+\hat v_1 q_1)^2\right)
\widetilde\Phi_{D}\left(-\,(k_2+\hat u_4 q_2)^2\right)
\nn
&\times& \Big\{
\,\,\Tr\left[\gamma_5   S_1(k_1) \gamma_5    S_3(k_1+q_2) \right] 
    \Tr\left[\gamma^\mu S_4(k_2) \gamma^\alpha S_2(k_2+q_1) \right] 
\Big\}
\nn[2ex]
&=& A_{D^\ast D}\,g^{\mu\alpha} + B_{D^\ast D}\,q_1^\mu q_2^\alpha\,.
\label{eq:decay-mol-DD2}
\ena
The argument of the $Z_c$-vertex function is given by
\bea
\vec\delta^{\,2} &=& \delta_1^2+\delta_2^2+\delta_3^2,
\nn
\delta_1 &=& - \frac{1}{2\sqrt{2}}
\left(k_1+k_2+(1+w_1-w_2) q_1+ (w_1-w_2)q_2) \right),
\nn
\delta_2 &=& + \frac{1}{2\sqrt{2}}
\left(k_1+k_2-(w_3-w_4) q_1 +(1-w_3+w_4) q_2 \right),
\nn
\delta_3 &=& + \frac{1}{2}\left(-k_1+k_2+(1-w_1-w_2) q_1-(w_1+w_2)q_2) \right).
\label{eq:arg-mol-DD2}
\ena
The quark masses are specified as $m_1=m_2=m_c$,  $m_3=m_4=m_d=m_u$,
and the two-body reduced masses as
$\hat v_2=m_2/(m_2+m_4)$,  $\hat v_4=m_4/(m_2+m_4)$, 
$\hat u_1=m_1/(m_1+m_3)$, and $\hat u_3=m_3/(m_1+m_3)$. 

As a guide to adjust the parameter $\Lambda_{Z_c}$  we take the experimental
values for decay widths given in Ref.~\cite{Ablikim:2013xfr}.
If the parameter $\Lambda_{Z_c}$ is varied in the limits 
$\Lambda_{Z_c}=3.3\pm 0.1$~GeV the numerical values of decay widths
vary according to 
\bea
\Gamma(Z^+_c\to\Jpsi+\pi^+) &=& (1.8 \pm 0.3)\,\text{MeV}\,,
\nn
\Gamma(Z^+_c\to\eta_c+\rho^+) &=& (3.2^{+0.5}_{-0.4})\,\text{MeV}\,,
\nn
\Gamma(Z^+_c\to \bar D^0 + D^{\ast\, +}) &=& (10.0^{+1.7}_{-1.4})\,\text{MeV}\,,
\nn
\Gamma(Z^+_c\to \bar D^{\ast\, 0} + D^+) &=& (9.0^{+1.6}_{-1.3})\,\text{MeV}\,.
\label{eq:Zc-mol-values}
\ena

Thus a molecular-type current for the vertex function of the $Z_c$ 
is in accordance with the experimental observation \cite{Ablikim:2013xfr} 
that $Z_c(3900)$ has a much stronger coupling to $DD^\ast$ than to
$\Jpsi\pi$.

\section{$X_b$ as a tetraquark}

Let us first interpret $X_b$ as a tetraquark state with the quantum numbers
$J^P=0^+$. Then the interpolating current for the
$X_b(5568)$ is given by 
\be
J = \varepsilon_{abc}\varepsilon_{dec}
(u^T_a C\gamma_5 b_b)(\bar d_d\gamma_5 C \bar s^T_e)
\label{eq:Xb-tet-cur}
\en 
The nonlocal version of the four-quark interpolating current reads

\bea
J^+_{X_b}(x) &=& \int\! dx_1\ldots \int\! dx_4 
\delta\left(x-\sum\limits_{i=1}^4 w_i x_i\right) 
\Phi_{X_b}\Big(\sum\limits_{i<j} (x_i-x_j)^2 \Big)
J^+_{4q}(x_1,\ldots,x_4),
\label{eq:Xb-nonlocal-cur}\\
J^+_{4q}&=& \varepsilon_{abc}\varepsilon_{dec} \,
[u_a(x_3)C\gamma_5 b_b(x_1)][\bar d_d(x_4)\gamma_5 C \bar s_e(x_2)].
\nonumber
\ena 
where $w_i=m_i/\sum_{j=1}^4 m_j$.
The effective interaction Lagrangian describing the coupling
of the meson $X_b$ to its constituent quarks takes the form
\be
{\cal L}_{\rm int} = g_{X_b}\,X^-_b(x)\cdot J^+_{X_b}(x) + \text{H.c.}
\label{eq:Xb-lag}
\en     

The Fourier transform of the $X_b$-tetraquark mass operator are given by
\bea
\Pi_{X_b}(p^2) &=& 
6\,\prod\limits_{i=1}^3\int\!\!\frac{d^4k_i}{(2\pi)^4i}\,
\widetilde\Phi_{X_b}^2\left(-\,\vec\omega^{\,2}\right) 
\nn
&\times& 
\,\,\Tr\left[\gamma_5 S_1(\hat k_1)\gamma_5 S_3(\hat k_3) \right]
    \Tr\left[\gamma_5 S_2(\hat k_2)\gamma_5 S_4(\hat k_4) \right] 
\label{eq:Xb-Fourier-mass}
\ena
where $\hat k_1=k_1-w_1 p$, $\hat k_2=k_2-w_2 p$,  $\hat k_3=k_3+w_3 p$,
      $\hat k_4=k_1+k_2-k_3+w_4 p$, and 
$\vec\omega^{\,2}=1/2\,(k_1^2+k_2^2+k_3^2+k_1k_2-k_1k_3-k_2k_3)$.
 
The matrix element of the decay  $X_b^+(p)\to B_s(q_1)+\pi^+(q_2)$ reads 
\bea
M\left( Z_b\to B_s+\pi^+\right)
&=&
6\,g_{X_b}g_{B_s}g_{\pi}
\nn
&\times&
\int\!\!\frac{d^4k_1}{(2\pi)^4i}\,\int\!\!\frac{d^4k_2}{(2\pi)^4i}\,
\widetilde\Phi_{X_b}\left(-\,\vec\eta^{\,2}\right)
\widetilde\Phi_{B_s}\left(-\,(k_1+v_1 q_1)^2\right)
\widetilde\Phi_{\pi}\left(-\,(k_2+u_4 q_2)^2\right)
\nn
&\times& 
Tr\left[ \gamma_5 S_1(k_1)\gamma_5 S_2(k_1+q_1) 
         \gamma_5 S_4(k_2)\gamma_5 S_3(k_2+q_2) \right] 
\nn[2ex]
&=& G_{X_bB_s\pi}\,.
\label{eq:Xb-decay}
\ena
where the arguments of the $X_b$-vertex function are given by
\bea
\vec\eta^{\,2} &=& \eta_1^2+\eta_2^2+\eta_3^2,
\nn
\eta_1 &=& - \frac{1}{2\sqrt{2}}\left(2k_1+(1+w_1-w_2) q_1 +(w_1-w_2) q_2 \right)\,,
\nn
\eta_2 &=& + \frac{1}{2\sqrt{2}}\left(2k_2-(w_3-w_4) q_1 +(1-w_3+w_4) q_2 \right)\,,
\nn
\eta_3 &=& + \frac{1}{2}\left((w_3+w_4) q_1 - (w_1+w_2) q_2 \right)\,.
\label{eq:Xb-arg-CC}
\ena
The quark masses are specified as $m_1=m_b$, $m_2=m_s$,  $m_3=m_u$, $m_4=m_d$,
and the two-body reduced masses as $v_1=m_1/(m_1+m_2)$,  $v_2=m_2/(m_1+m_2)$, 
$u_3=m_3/(m_3+m_4)$, and $u_4=m_4/(m_3+m_4)$. 

The two-body decay width is given by
\be
\Gamma(X_b\to B_s+\pi) =
\frac{\bf|q_1|}{8\pi M_{X_b}^2}\,G^2_{X_bB_s\pi}\,,
\label{eq:Xb-width}
\en
where ${\bf|q_1|}$ 
is the momentum of the daughter particles in the rest frame of the $X_b$. 

We adjust the parameter $\Lambda_{X_b}$ for two values of 
the $X_b$ mass, (i) $m_{X_b}$=5567.8~MeV as reported by
the D0 Collaboration~\cite{D0:2016mwd},  
and 
(ii)  $m_{X_b}$=5771~MeV as was obtained in  \cite{Ali:2016gdg}.
The numerical values of the decay widths can be calculated to be     
\bea
m_{X_b}&=& 5.568~\text{GeV}, \quad \Lambda_{X_b} = (1.36\pm 0.05)~\text{GeV}\,,
\quad  \Gamma(X_b\to B_s\pi) = (21.9 \pm 3.5)~\text{MeV}\,,
\nn
m_{X_b}&=& 5.771~\text{GeV}, \quad \Lambda_{X_b} = (1.66\pm 0.05)~\text{GeV}\,,
\quad  \Gamma(X_b\to B_s\pi) = (21.7 \pm 3.5)~\text{MeV}\,,
\label{eq:Xb-tetra-values}
\ena 

\section{$Z(4430)$ as a tetraquark}

The interpolating tetraquark current of the $Z(4430)$ state with $J^P = 1^+$ fixed
by the LHCb Collaboration~\cite{4430lhcb1} has the same structure
as the tetraquark current for the $Z_c$ state
[see Eqs.~(\ref{eq:Zc-tet-cur}) and~(\ref{eq:nonlocal-cur})].
Similarity of the $Z(4430)$ and $Z_c$ states also concerns 
the effective interaction Lagrangian describing the coupling of $Z(4430)$
to its constituent quarks, 
\bea
{\cal L}_{\rm int} = g_Z Z_\mu(x) \cdot J^\mu_Z(x) + {\rm H.c.}\,,
\ena
where $J^\mu_Z(x) = J^\mu_{Z_c}(x)$ with a specific value of the size parameter $\Lambda_Z$.

In the case of $Z(4430)$ we consider two strong decay modes
$Z(4430) \to J/\psi + \pi$ and $Z(4430) \to \psi(2s) + \pi$, which are calculated by
analogy with the case of $Z_c \to J/\psi + \pi$ in the tetraquark picture. 
A new feature is that we should specify
the vertex function of the $\psi(2s)$ state. By analogy with the oscillator potential model
it should emulate the node structure of the $\psi(2s)$. In our calculations we use
the following form of the $\psi(2s)$--vertex function:
\bea
\tilde\Phi_{\psi(2s)}(-k^2) = \exp(k^2/\Lambda_{\psi(2s)}) \,
\biggl[ 1 - \alpha
\exp(k^2/\Lambda_{\psi(2s)}) \biggr] \,,
\ena
where $\alpha$ is a free parameter, encoding the node structure of the
$\psi(2s)$ meson. It is fixed at $\alpha = 1.0172$ from 
the description of the leptonic decay constant $f_{\psi(2s)} = 291$ MeV. 
For convenience, we use the same size parameter
$\Lambda_{J/\psi} = \Lambda_{\psi(2s)} = 1.738$ GeV for
$J/\psi$ and its radial excitation $\psi(2s)$ state.
For the $Z(4430)$ mass we use the actual value 4.478 GeV.

Now let us turn to the discussion of our results for the
$Z(4430) \to J/\psi + \pi$ and $Z(4430) \to \psi(2s) + \pi$ decay widths.
We have a single free parameter: the $\Lambda_{Z(4430)}$ - size parameter
of the $Z(4430)$ state.
We use the present upper limit for the total width of the $Z(4430)$ state
$\Gamma \le 212$ MeV deduced from the averaged value $\Gamma = 181 \pm 31$ MeV
in Particle Data Group~\cite{Agashe:2014kda} as the upper limit
for the sum of the widths of
two modes $Z(4430) \to J/\psi + \pi$ and $Z(4430) \to \psi(2s) + \pi$.
It constrains the choose of the size parameter $\Lambda_Z$.
In particular, we found that $\Lambda_Z \ge 2.2$ GeV,
which supports the compact $(c \bar c d \bar u)$ tetraquark interpretation of the
$Z(4430)$ state.
In Table~\ref{tab:Xcs0} we present our numerical results for the partial decay widths
$\Gamma_{J/\psi} \doteq \Gamma(Z(4430) \to J/\psi + \pi)$, and
$\Gamma_{\psi(2s)} \doteq \Gamma(Z(4430) \to \psi(2s) + \pi)$ decay widths,
their sum $\Gamma = \Gamma_{J/\psi} + \Gamma_{\psi(2s)}$
and their ratio $R_Z = \Gamma_{\psi(2s)}/\Gamma_{J/\psi}$
for variation of $\Lambda_Z$ from 2.2 to 3.2 GeV.
One can see that the decay width of $Z(4430) \to \psi(2s) + \pi$ process
dominates over the one of the $Z(4430) \to J/\psi + \pi$
by a factor $R_Z \simeq (4.36 \pm 0.28)$.

\begin{table}[hb]
\begin{center}
\caption{$Z(4430)$ decay rates.}
\def\arraystretch{1.25}
\begin{tabular}{|c|c|c|c|c|}
\hline
$\Lambda_{Z(4430)}$ (GeV) & $\Gamma_{J/\psi}$ (MeV) & $\Gamma_{\psi(2s)}$ (MeV)
& $\Gamma$ (MeV) & $R_Z$ \\
\hline
2.2 & 37.4 & 173.7 & 211.1 & 4.64 \\
2.3 & 31.7 & 144.7 & 176.4 & 4.56 \\
2.4 & 26.9 & 120.6 & 147.5 & 4.48 \\
2.5 & 22.9 & 100.8 & 123.7 & 4.40 \\
2.6 & 19.4 &  84.4 & 103.8 & 4.35 \\
2.7 & 16.5 &  70.9 &  87.4 & 4.30 \\
2.8 & 14.1 &  59.7 &  73.8 & 4.23 \\
2.9 & 12.0 &  50.4 &  62.4 & 4.20 \\
3.0 & 10.3 &  42.7 &  53.0 & 4.15 \\
3.1 &  8.8 &  36.3 &  45.1 & 4.13 \\
3.2 &  7.6 &  31.0 &  38.6 & 4.08 \\
\hline
\end{tabular}
\label{tab:Xcs0}
\end{center}
\end{table}

Finally, we make the prediction for the 
$Z(4430)^+ \to D^{\ast\,+} + \bar D^{\ast\,0}$ decay rate. 
This process is described by the invariant matrix element, 
which is expressed in terms of three relativistic amplitudes 
$B_i$, $(i=1,2,3)$ as  
\bea 
M^{\mu\alpha\beta}(Z(4430)(p,\mu) \to 
D^{\ast}(q_1,\alpha) + \bar D^{\ast}(q_2,\beta)) = 
  B_1 q^\mu_1 \epsilon^{q_1q_2\alpha\beta} 
+ B_2 \epsilon^{q_1\mu\alpha\beta} 
+ B_3 \epsilon^{q_2\mu\alpha\beta} \,.
\ena 
The $Z(4430)^+ \to D^{\ast\,+} + \bar D^{\ast\,0}$ decay rate   
is calculated according to the formula 
\bea 
\Gamma(Z(4430)^+ \to D^{\ast +} + \bar D^{\ast 0}) &=&
\frac{\bf|q_1|}{12 \pi M_{Z}^2}\nonumber\\
&\times& 
\biggl[ B_1^2 M_Z^2 {\bf|q_1|}^4 
+  B_2^2 \biggl( 3 M_{D^{\ast +}}^2 + 
\Big( 1+ \frac{M_Z^2}{M_{D^{\ast 0}}^2}\Big) {\bf|q_1|}^2 \biggr) \nonumber\\
&+&  B_3^2 \biggl( 3 M_{D^{\ast 0}}^2 + 
\Big( 1+ \frac{M_Z^2}{M_{D^{\ast +}}^2}\Big) {\bf|q_1|}^2 \biggr) \nonumber\\
&+& B_1 B_2 {\bf|q_1|}^2 \Big( M_{Z}^2 + M_{D^{\ast +}}^2 - 
M_{D^{\ast 0}}^2 \Big) \nonumber\\
&+& B_1 B_3 {\bf|q_1|}^2 \Big( M_{Z}^2 + M_{D^{\ast 0}}^2 - 
M_{D^{\ast +}}^2 \Big) \nonumber\\  
&+& B_2 B_3 \Big( 3 (M_Z^2 - M_{D^\ast +}^2 - M_{D^{\ast 0}}^2) - 
{\bf|q_1|}^2 \Big)
\biggr]\,. 
\ena 
Our numerical result for $\Lambda_{Z(4430)}$ varied from 2.2 to 3.2 GeV 
is $\Gamma(Z(4430)^+ \to D^{\ast +} + \bar D^{\ast 0}) = 23.5 \pm 15.6$ MeV. 

\section{Summary and conclusions}

Let us summarize the main results of our paper. Presently 
two possible four-quark configurations for exotic states are tested 
experimentally and theoretically: the tetraquark (compact) configuration 
corresponding to the coupling of color diquark and antidiquark and 
molecular (extended) configuration corresponding to the coupling 
of two separate mesons. 
We have critically checked both possible four-quark pictures 
(tetraquark and molecular scenario) in the case of the $Z_c(3900)$ state. 
For the case of the $X(5568)$ and $Z(4430)$ states we considered only 
the tetraquark picture. Our study has been done 
by analyzing strong decays of the exotic state. 
The strong decays have been calculated in the framework of the covariant
quark model previously developed by us.
First, we have interpreted the  $Z_c(3900)$ state 
as the isospin 1 partner of the $X(3872)$.
We have calculated the partial widths of the decays
$Z_c^+(3900)\to\Jpsi\pi^+$, $\eta_c\rho^+$, and $\bar D^0D^{\ast\,+}$, 
$\bar D^{\ast\,0}D^{+}$. 
It turned out that the leading metric Lorentz structure in the matrix elements
describing the decays $Z_c(3900)\to\bar DD^{\ast}$ vanishes analytically.
This results in a significant $D$--wave suppression of these decays through the
appearance of the phase space factor proportional to $|{\bf q}|^5$.
Since the experimental data from the BESIII Collaboration show
that $Z_c(3900)$ has a much more stronger coupling to $DD^\ast$ than to
$\Jpsi\pi$, we  have concluded that the tetraquark-type current  
for the $Z_c(3900)$ is in disaccord with experiment.
As an alternative we have employed a molecular-type four-quark current to
describe the $Z_c(3900)$ state. In this case  we found that for a
relatively large model size parameter of $\Lambda_{Z_c} \sim 3.3$~GeV
one can obtain partial widths for the decays $Z_c(3900)\to\bar DD^{\ast}$ 
that are close to $\sim 15$~MeV for each mode. At the same time
the partial widths for the decays $Z_c(3900)\to\Jpsi\pi\,, \eta_c\rho$
are suppressed by a factor of $6-7$ in accordance with experimental
data.  

Finally, we have tested a tetraquark picture for the 
$X(5568)$ and $Z(4430)$ structure by analyzing their strong one-pion decay.
In the analysis of the $B_s \pi$ decay mode of the $X(5568)$ 
we found that one can fit the experimental 
decay width using  a mass of $5568$~MeV by taking the value of the 
parameter to be
$\Lambda_{X_b}\sim 1.4$~GeV. In the case of a larger mass  
$5771$~MeV one finds  $\Lambda_{X_b}\sim 1.7$~GeV. 
In the case of the $Z(4430)$ state we considered the modes with $J/\psi$ and
its first radial excitation $\psi(2s)$.
We showed that the decay width of the $Z(4430) \to \psi(2s) + \pi$ process dominates
over the one of $Z(4430) \to J/\psi + \pi$ by a factor $R_Z = (4.36 \pm 0.28)$
and the sum of the two decay rates of the $Z(4430)$ satisfies the upper limit for 
the total width of the $Z(4430)$ if the size parameter $\Lambda_{Z(4430)} \ge 2.2$ GeV.
It means that the $Z(4430)$ state is a good candidate for the compact tetraquark state. 
Our prediction for the $Z(4430)^+ \to D^{\ast +} + \bar D^{\ast 0}$ decay width is 
$\Gamma(Z(4430)^+ \to D^{\ast +} + \bar D^{\ast 0}) = 23.5 \pm 15.6$ MeV. 

\begin{acknowledgments}

This work was supported
by the German Bundesministerium f\"ur Bildung und Forschung (BMBF)
under Project 05P2015 - ALICE at High Rate (BMBF-FSP 202):
``Jet- and fragmentation processes at ALICE and the parton structure 
of nuclei and structure of heavy hadrons'', 
by Tomsk State University Competitiveness 
Improvement Program and the Russian Federation program ``Nauka'' 
(Contract No. 0.1526.2015, 3854). 
M.A.I.\ acknowledges the support from the PRISMA cluster of excellence 
(Mainz University). M.A.I. and J.G.K. thank the Heisenberg-Landau grant for
partial support.  

\end{acknowledgments}

\begin{appendix}
  \section{Spin kinematics for the decay $1^+ \to 1^- +0^-$}
The matrix element 
\be
M=\langle 1^- (q_1;\rho),0^-(q_2)|\,T\,|1^+(p;\mu)\rangle
\label{eq:A1}
\en
can be described by the three sets of amplitudes: (i) invariant amplitudes, 
(ii) helicity amplitudes, and (iii) $(LS)$ amplitudes. In this Appendix we
derive the relations between the three sets of amplitudes.

The product of the parities of the two final state mesons is $(+1)$ which
    matches the parity of the initial state. Thus the two final state mesons
    must have even relative orbital momenta. In
    the present case these are $L=0,2$. The spins $s_1$ and $s_2$ of the two
    final state mesons couple to the total spin
  $S=1$. Thus one has the two $(LS)$ amplitudes
\be
A_{01}\, ,  \quad A_{21}
\label{eq:A2}  
\en

  There are two covariants ${\cal K}_1^{\mu \rho}=m\, g^{\mu\rho}$ and
  ${\cal K}_2^{\mu \rho}=\frac{1}{m}\,q_1^\mu\,q_2^\rho$ that describe the
  matrix element. These define the invariant amplitudes $A_1$ and $A_2$
  according to
\be
M\,=\,(A_1\,{\cal K}_1^{\mu \rho}+ A_2\,{\cal K}_2^{\mu\rho})\,
\varepsilon_\mu\,\varepsilon^\ast_{1\rho}
\label{eq:A3}
\en
  There are two independent helicity amplitudes $H_{\lambda\,\lambda_1}$
$(\lambda=\lambda_1)$\,, 
  \be
  H_{+1\,+1} \,, \quad H_{0\,0}
\label{eq:A4}
  \en
  From parity one has $H_{-1\,-1}=H_{+1\,+1}$.
  In order to relate the helicity amplitudes to the invariant amplitudes we
work in the rest system of the decay meson and define the
$z$ direction to be along the momentum direction of meson 1.
  The helicity amplitudes can be related to the invariant amplitudes using
  the momenta and polarization vectors 
  $\epsilon_1^\rho(\pm)=
(0;\mp 1,-i,0)/\sqrt{2},\, 
\,\epsilon_1^\rho(0) =
(\,|{\bf q_1}|\,;\,0\,,\,0\,,\,E_1\,)/m_1,\,\, 
\epsilon^\mu(\pm)=
(0;\mp 1,-i,0)/\sqrt{2}, 
\epsilon^\mu(0) = (0;0,0,1),\, 
q_1^\mu=(E_1;0,0,|{\bf q_1}|)\,, \quad q_2^\mu=(E_2;0,0,-|{\bf q_1}|)$.  
One can then express the helicity amplitudes in terms of the invariant
amplitudes. The relations can be calculated to be
\bea
H_{00}&=&-\,\frac{m}{m_1}E_1\,A_1
-\frac{1}{m_1}|{\bf q_1}|^2 \,A_2\nn
H_{+1+1}&=&H_{-1-1}=-\,m\,A_1
\label{eq:A5}
\ena
where the magnitude of the final state three-momentum is given by
$|{\bf q_1}|=\sqrt{Q_+ Q_-}/2m$ with $Q_\pm = m^2-(m_1 \pm m_2)^2=
2(q_1q_2 \mp m_1m_2)$.


The coefficients of the matrix relating the $(LS)$ and helicity amplitudes
can be calculated from the product of two C.G. coefficients according to
\cite{Martin}
\be
\langle JM;LS|JM;\lambda_1 \lambda_2 \rangle =
\left(\frac{2L+1}{2J+1}\right)^{1/2}
\langle LS;0\mu|J\mu \rangle
\langle s_1 s_2;\lambda_1,-\lambda_2|S\mu \rangle 
\label{LShel}
\en
where $\mu=\lambda_1-\lambda_2$. One obtains
  \be
  \left(
\begin{array}{c}
  A_{01} \\[0.5ex]
A_{21}
\\[0.0ex] 
\end{array}
\right)
=\sqrt{\frac{1}{3}}\left(
\begin{array}{cc}
  2 & 1 \\[0.5ex]
\sqrt{2} &-\sqrt{2}\\[0.0ex] 
\end{array}
\right)
\left(
\begin{array}{c}
  H_{+1+1} \\[0.5ex]
 H_{00}
\\[0.0ex] 
\end{array}
\right)
\label{eq:A7}
\en 
We can thus relate the $(LS)$ amplitudes to the invariant amplitudes $A_i$.
The relations read
\bea
A_{01}&=&-\sqrt{\frac{1}{3}}\frac{1}{m_1}
\left(m(2m_1+E_1)A_1 +|{\bf q_1}|^2 A_2 \right) \nn
A_{21}&=&\sqrt{\frac{2}{3}}\frac{1}{m_1}
\left(m(E_1-m_1)A_1 +|{\bf q_1}|^2 A_2 \right)
\label{eq:A8}
\ena
The $(LS)$ amplitude $A_{21}$ can be seen to have the correct  $D$--wave
threshold behavior proportional to $|\vec q_1|^2$ by taking the relation
$(E_1-m_1)=|{\bf q_1}|^2/(E_1+m_1)$ into account.

The rate for the decay process $1^+(p) \to 1^-(q_1)+0^-(q_2)$ is given by
  \bea
  \Gamma&=&\frac{1}{8\pi}\frac{1}{2s+1}\frac{|{\bf q_1}|}{m^2}
  (|H_{+1+1}|^2+|H_{-1-1}|^2 +|H_{00}|^2) \nn
&=&\frac{1}{8\pi}\frac{1}{2s+1}\frac{|{\bf q_1}|}{m^2}
  (|A_{01}|^2 +|A_{21}|^2)
\label{eq:A9}
    \ena
  where $2s+1=3$.

  We assume that the $(1^-)$ meson decays into two pseudoscalar mesons as
  in the cascade decay $Z_c \to D +D^\ast(\to D+\pi)$. We treat the cascade
  decay in the narrow width approximation. The differential decay
  distribution for the cascade decay is given by
  \be
  \frac{d\Gamma(Z_c \to D +D^\ast(\to D+\pi))}{d\cos\theta}=B(D^\ast\to D+\pi)
  \frac{1}{24\pi}\frac{|{\bf q_1}|}{m^2}\Big(
  \frac 38\,(1+\cos^2\theta) {\cal H}_T
  + \frac 34 \sin^2\theta\,{\cal H}_L\Big)
\label{eq:A10}
  \en
  where ${\cal H}_T\,=\,|H_{+1+1}|^2+|H_{-1-1}|^2 , {\cal H}_L\,=\,|H_{00}|^2$.
  For the cascade decay $Z_c \to \pi + J/\psi (\to\ell^+ \ell^-)$ we again have
\be
  \frac{d\Gamma(Z_c \to \pi + J/\psi (\to\ell^+ \ell^-))}
       {d\cos\theta}=B(J/\psi \to\ell^+ \ell^-)
  \frac{1}{24\pi}\frac{|{\bf q_1}|}{m^2}\Big(
  \frac 38\,(1+\cos^2\theta) {\cal H}_T
  + \frac 34 \sin^2\theta\,{\cal H}_L\Big)
 \label{eq:A11}
 \en
  In the latter cascade decay we have set $m_\ell=0$.

 \end{appendix}

\end{document}